\documentclass{emulateapj}

\newcommand{\um}{$\mu$m}

\newcommand{\gap}{$\gtrsim$}

\newcommand{\sqam}{arcmin$^2$}
\newcommand{\s}{$\sigma$}
\newcommand{\persqdeg}{deg$^{-2}$}

\slugcomment{ApJL, in press}

\shorttitle{SCUBA Map in the {\it Spitzer} FLS}
\shortauthors{Sawicki \& Webb}

\begin{document}

%% LaTeX will automatically break titles if they run longer than
%% one line. However, you may use \\ to force a line break if
%% you desire.

\title{A SCUBA Map in the {\it Spitzer} First Look Survey: 
Source Catalog and Number Counts}

\author{
Marcin Sawicki} 
\affil{
Dominion Astrophysical Observatory, 
Herzberg Institute of Astrophysics,
National Research Council, 
5071~West Saanich Road, 
Victoria, B.C., V9E 2E7, 
Canada
}
\email{marcin.sawicki@nrc.gc.ca}

\author{\vspace{-5mm}and\\
T.M.A.\ Webb} 
\affil{Sterrewacht Leiden, 
Niels Bohrweg 2, 
Leiden NL-2333 CA, 
The Netherlands
}
\email{webb@strw.leidenuniv.nl}

\begin{abstract}
Using the SCUBA instrument on the JCMT, we have made a submillimeter
mosaic at 850\um\ of a subarea of the {\it Spitzer} First Look Survey
(FLS).  Our image covers the central 151 \sqam\ of the northern
extragalactic Continuous Viewing Zone (CVZ) field of the FLS to a
median 3\s\ depth of 9.7 mJy.  The image contains ten 850\um\ sources
detected at 3.5\s\ or higher significance, of which five are detected
at $\geq$4\s.  We make the catalog of these SCUBA-selected FLS sources
available to the community.  After correcting for incompleteness and
flux bias, we find that the density of sources brighter that 10mJy in
our field is (1.3$^{+1.1}_{-0.7}$)$\times$$10^2$ \persqdeg\ (95\%
Poisson confidence limits), which is consistent with other surveys
that probe the bright end of the submillimeter population.

\end{abstract}


\section{INTRODUCTION}

Launched in the second half of 2003, the Spitzer Space Telescope
(formerly SIRTF), the fourth and final of NASA's Great Observatories,
holds the promise of addressing many outstanding questions related to
dust-enshrouded galaxy formation at high redshift.  One of the first
science observations to be undertaken by {\it Spitzer} is the {\it
Spitzer} First Look Survey (FLS), a first look at the mid-IR sky at
sensitivities that are two orders of magnitude deeper than previous
large-area surveys.  In addition to {\it Spitzer} data at 3.6, 4.5,
5.8, 8.0, 24, 70, and 160\um, the FLS has also been observed in deep
ground-based campaigns at optical (KPNO 4m to $R$=25.5, 5\s\ in a
2\arcsec\ aperture) and radio (VLA to 115$\mu$Jy, 5\s, per 5\arcsec\
beam at 1.4GHz; Condon et al.\ 2003) wavelengths\footnote{See
http://ssc.spitzer.caltech.edu/fls/}.  The {\it Spitzer} FLS data will
be released to the public in early 2004 and, together with the deep
ancillary ground-based data, will provide the community with the first
systematic look at the properties of faint {\it Spitzer}-selected
extragalactic sources.

While {\it Spitzer} will discover many dust-enshrouded high-$z$
objects and will greatly help us understand their nature, the fact
that it does not image at wavelengths longward of 160\um\ presents
some important limitations.  For example, for typical dust
temperatures (20--40K; e.g., Dunne et al.\ 2000), the longest {\it
Spitzer} passband barely probes longward of the peak of the thermal
dust emission for galaxies at even moderate redshifts, making it
difficult to estimate their bolometric luminosities and hence infer
quantities such as dust masses and star formation rates.  Moreover,
with increasing redshift (or decreasing dust temperature), an object's
dust emission peak shifts redward of the longest {\it Spitzer}
wavelength, causing strong negative k-corrections and making a galaxy
at high redshift (or one with low dust temperatures) fade rapidly out
of a Spitzer-selected sample.

In this Letter we present complementary long-wavelength imaging
observations of a section of the {\it Spitzer} FLS, obtained at
850\um\ with the Submillimetre \notetoeditor{Editor: this is an
instrument whose follows the British spelling convention of
submillimetre/submillimeter} Common User Bolometer Array (SCUBA) on
the James Clerk Maxwell Telescope (JCMT).  In a future paper we will
discuss in detail the multiwavelength properties of sources in the
area of our SCUBA map; the purpose of the present Letter is to quickly
make available to the community the source catalog of objects from our
SCUBA observations within the public-release {\it Spitzer} FLS.

\section{THE DATA}

\begin{figure}
\epsscale{1.0}
\plotone{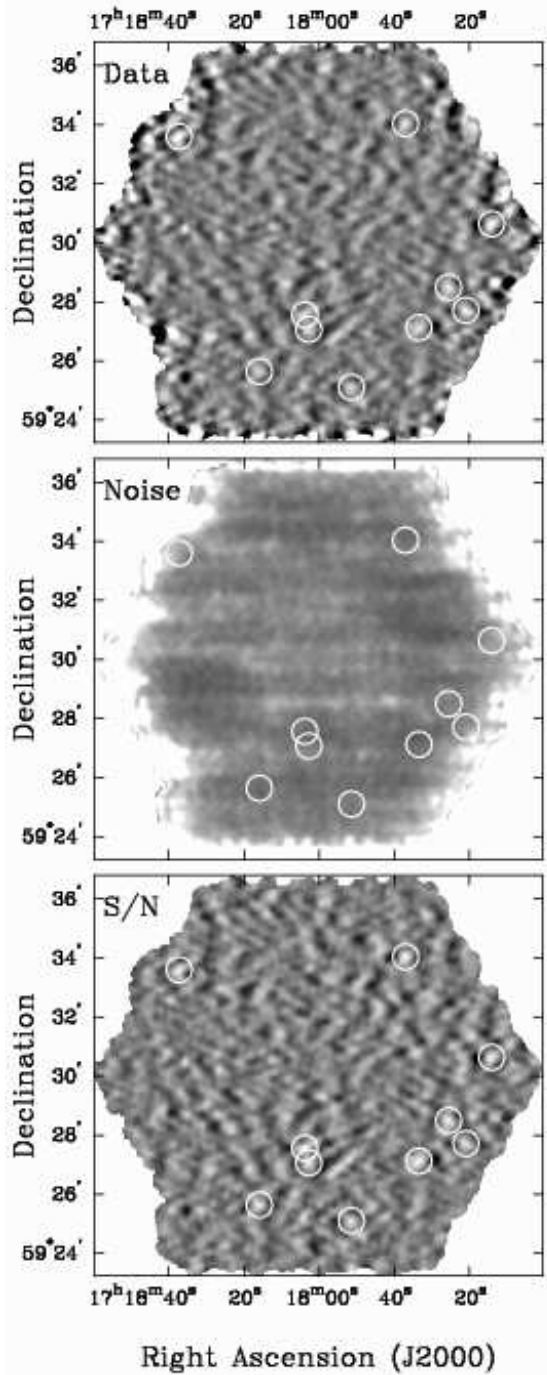}
\caption{\label{maps.fig} The beam-convolved signal (top), noise
(middle), and RMS signal-to-noise (bottom) maps of our 850\um\ survey.
These maps are mosaics of multiple individual SCUBA pointings. 
Sources detected at $>$3.5\s\ are marked with circles.  A HIGH QUALITY VERSION OF THIS FIGURE CAN BE OBTAINED FROM THE AUTHORS.}
\end{figure}

\subsection{Observations}

We used the SCUBA instrument (Holland et al.\ 1999) on the JCMT to
observe a contiguous area of 151 \sqam\ centered on the nominal center
of the FLS northern extragalactic CVZ field at R.A.=17:18:00,
Dec=+52:24:30 (J2000.0).  SCUBA is an array of 37 bolometers at
850$\mu$m and 91 at 450$\mu$m that is able to observe simultaneously
at both wavelengths, although particularly excellent weather is
required for 450$\mu$m observations.  In its jiggle-map mode --- which
is the mode we used --- a SCUBA observation has a footprint of
$\sim$2\arcmin\ diameter; to cover a large contiguous area we tiled
our field in a spiral pattern on a hexagonal grid, starting at the
nominal central FLS coordinates.  A typical point in our combined map
received a total of 2048 seconds of integration, split into 4 visits
that were separated in time by many hours and, often, nights.

The data were obtained during thirteen nights from 2002 March through
2003 March, using a total of 7.5 usable (out of 10 allocated) shifts
of JCMT time.  The weather varied from grade 1 to grade 4, or
${\tau}_{850}\sim$ 0.1 to ${\tau}_{850}\sim$ 0.45, where
${\tau}_{850}$ is a measure of the optical depth at 850$\mu$m.  To
remove the rapidly-varying sub-mm sky we used the standard SCUBA
chopping technique with a {\it chop-throw} of 30\arcsec\ held constant
in RA. This technique produces the familiar {\it
negative-positive-negative} beam pattern apparent in many SCUBA maps
and can be used to increase the significance of detection for
individual sources by taking advantage of the signal in the off-beams.

Throughout the observing campaign sky opacity was measured using
skydip observations every $\sim$1.5 hours, though less often in
exceptionally transparent and stable weather. Pointing checks were
performed every 1.5 hours and the data were flux calibrated using
standard JCMT flux calibrators.

\begin{deluxetable}{lrrrr}
%\tabletypesize{\scriptsize}
%\tablewidth{0 in}
\tablecaption{\label{sources.tab} 850$\mu m$ sources detected at $>$3.5$\sigma$. \hspace{25mm}}
\tablehead{
\colhead{ID} &              
\colhead{RA(2000)} & 
\colhead{Dec(2000)} & 
\colhead{S$_{850}$(mJy)} & 
\colhead{S/N} \\
}
\startdata
FLS850.1803+2733  & 17:18:03.9 &  +59:27:33 &  10.9$\pm$2.4 &   4.5 \\
FLS850.1736+3401  & 17:17:36.9 &  +59:34:01 &  11.3$\pm$2.6 &   4.4 \\
FLS850.1733+2706  & 17:17:33.5 &  +59:27:06 &  11.0$\pm$2.6 &   4.3 \\
FLS850.1725+2828  & 17:17:25.6 &  +59:28:28 &  12.3$\pm$3.0 &   4.2 \\
FLS850.1837+3335  & 17:18:37.3 &  +59:33:35 &  16.0$\pm$4.0 &   4.0 \\
FLS850.1714+3036  & 17:17:14.1 &  +59:30:36 &  14.6$\pm$3.7 &   3.9 \\
FLS850.1816+2538  & 17:18:16.0 &  +59:25:38 &   9.5$\pm$2.4 &   3.9 \\
FLS850.1751+2505  & 17:17:51.5 &  +59:25:05 &  10.5$\pm$2.8 &   3.8 \\
FLS850.1721+2741  & 17:17:21.0 &  +59:27:41 &  13.5$\pm$3.7 &   3.7 \\
FLS850.1802+2703  & 17:18:02.8 &  +59:27:03 &  10.7$\pm$2.9 &   3.7 \\
\enddata
\end{deluxetable}

\subsection{Data reduction}

The data were reduced using the standard SCUBA User Reduction Facility
(SURF) procedure.  After removing the {\it nod}, the data were
flatfielded and corrected for sky opacity using the skydip
measurements.  At each second the mean sky level was subtracted from
the array and noise spikes were iteratively removed from the bolometer
timestreams.  Finally, the data were rebinned onto the sky plane to
produce the final (unsmoothed) map.

To increase the sensitivity to point sources, the unsmoothed map was
convolved with a template beam profile that was made from observations
of point-like calibration sources and contains the {\it
negative-positive-negative} beam pattern.  This technique reduces the
frequency of spurious sources which do not convolve well with the
beam, and increases the S/N by incorporating the flux from the two
off-source positions into the final flux measurements.  The top panel
of Figure~\ref{maps.fig} shows our beam-convolved map.

We used the method of Eales et al.\ (2000) to estimate the noise
across our map. We produced Monte Carlo simulations of each raw
bolometer time stream using the same noise level as in the real data
but adding no signal. These simulated data were then reduced using the
same set of steps as the real data resulting in a simulated sky
map. We produced 500 such simulated sky maps and the noise at each
pixel in the data map, shown in the middle panel of
Figure~\ref{maps.fig}, is taken to be the variance between these 500
simulations at that pixel.  The noise level determined by this method
agrees well with the noise estimated from the real data map.  The
median noise value is 3.2 mJy (1$\sigma$) and the spatial distribution
of the noise level is quite uniform across the entire field except
near its edges.

\subsection{Source detection and the object catalog}\label{catalog}

We used a combination of the data and noise maps to perform an
automated source search.  A S/N map (bottom panel of
Figure~\ref{maps.fig}) was produced by dividing the beam-convolved
data map by the beam-convolved noise map; all peaks with S/N$\geq$3.5
in this S/N map are source candidates.  To make the source finding
procedure objective and automated (a must for our completeness
simulations in \S~\ref{sourcecounts}), the actual search is performed
using the SExtractor source-detection software (Arnouts \& Bertin,
1996) on a truncated version of the S/N map.  Specifically, to
restrict SExtractor to sources with S/N$\geq$3 and to suppress
confusion due to the negative off-beams of bright sources, we set all
values S/N$<$3 to zero before running SExtractor.  As confirmed by
visual inspection, this technique reliably finds all the peaks above
S/N$>$3.  There are 27 peaks with S/N$>$3, and 10 with S/N$>$3.5.

\begin{figure}
\plotone{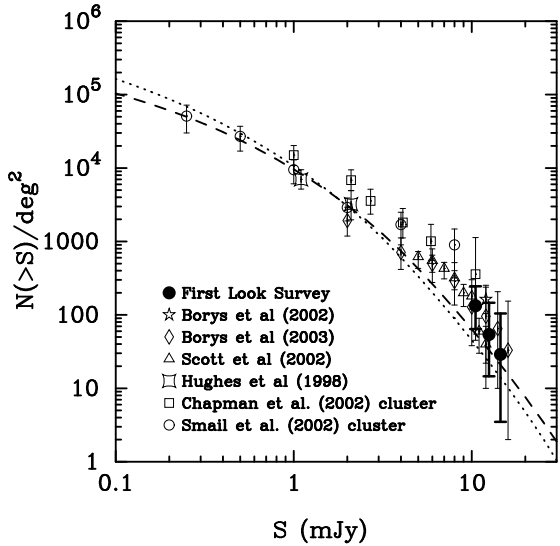}
\caption{\label{counts.fig} The cumulative 850\um\ number counts.  
The filled circle shows the results from the present work in the
Spitzer FLS.  Error bars are 95\% Poisson confidence limits.
Following Borys et al.\ (2003), overlaid are two predictions based on
representative galaxy evolution models from Rowan-Robinson (2001) ---
the dashed line is for a universe with
($\Omega_M,\Omega_\Lambda$)=(1.0, 0.0), and the dotted line is for
($\Omega_M,\Omega_\Lambda$)=(0.3, 0.7).  }
\end{figure}

Tables~\ref{sources.tab} presents all sources detected with
S/N$>$3.5. Column 1 gives the source ID, columns 2 and 3 give the
coordinates of the source determined by SExtractor in the truncated
S/N map, column 4 gives the source flux, and column 5 lists the
significance of the detection.

Based on Gaussian noise statistics we expect that at S/N=3.5 we should
have $\sim$1 spurious source in our map (the number of expected
spurious sources is close to zero at S/N=4).  Contamination by
spurious sources does not have a large effect on the sub-mm source
counts we discuss in the following section (\S~\ref{sourcecounts}),
but we mention it here to caution the reader that, statistically,
$\sim$1 of the sources in Table~\ref{sources.tab} may not be real.

\section{SOURCE COUNTS}\label{sourcecounts}

To date, two surveys have targeted the bright end ($S_{850\mu m}$\gap
10mJy) of the sub-mm population and both show a steep slope of the
cumulative source counts.  Scott et al.\ (2002) surveyed two spatially
independent fields with a total area of 260
\sqam, while Borys et al.\ (2002, 2003) studied an area of 165 \sqam\
in the region of the northern Hubble Deep Field.  Both surveys show
strong qualitative clustering of sources which may skew their source
count results.  Here, we use our 151 \sqam\ FLS SCUBA map to make a
third, independent measurement of the submillimeter source counts at
the bright end of the population.

\begin{deluxetable}{lccc}
%\tabletypesize{\scriptsize}
%\tablewidth{0 in}
\tablecaption{\label{FLScounts.tab} 850$\mu m$ number counts in the Spitzer FLS}
\tablehead{
\colhead{$S_{850}$(mJy)} &              
\colhead{$n_{objects}$} & 
\colhead{raw $N(>S)$} &
\colhead{corrected $N(>S)$} \\
}
\startdata
10 &	9 &	215 $^{+194}_{-112}$	&	133 $^{+112}_{-69}$ \\	
12 &	4 &	95  $^{+148}_{-69}$	&	54  $^{+92} _{-39} $ \\	
14 &	2 &	48  $^{+125}_{-42}$	&	29  $^{+76} _{-26} $ \\	
\enddata
\end{deluxetable}

Our {\it raw} cumulative source counts are presented in column 3 of
Table~\ref{FLScounts.tab}, where we have counted all objects {\it
detected} at S/N$>$3.5.  However, we are interested in sources that
are close to the noise level and so, to properly calculate the source
density, we must account for two effects: incompleteness and flux
bias.  Clearly, the number counts of faint sources near the detection
threshold will suffer from incompleteness, making it necessary to
correct their observed numbers upwards.  Additionally, however,
detected sources will also have suffered from flux boosting: while
sources whose flux densities are scattered below the detection
threshold are not counted in the sample, those that are scattered into
the sample from below the detection threshold will necessarily have
their flux densities overestimated.  These two effects compete against
each other, but for a source population where numbers increase quickly
with decreasing true flux density (as is the case here) flux boosting
should dominate.

We studied these issues using Monte Carlo simulations that implant
artificial sources into our data and seek to recover them using the
same technique that we used for constructing our source catalog in
\S~\ref{catalog}.  We generated artificial sources by flux-scaling the
empirical beam map constructed from observations of bright
point-source flux calibrators.  An artificial source was then added at
a random (but known) location to the data map, the resulting map was
divided by the noise map to form the S/N map, and then the
object-finding and flux measurement procedures used on the real data
were applied to search for the artificial object.  To statistically
assess the completeness and flux bias properties of our map, this
procedure was repeated, one artificial object at a time, for a range
of input fluxes and spatial positions.

This Monte Carlo procedure applied to our SCUBA map results in a
matrix, $S_{true,obs}$, that gives the probability that, for a source
of a known input flux density $S_{true}$, we will recover an object of
an observed flux density $S_{obs}$.  To understand the incompleteness
and flux bias effects on the {\it population} of sources, we need to
consider their effects on a plausible true source count distribution.
Following Borys et al.\ (2003) we assume the following functional form
to describe the underlying source count population:
\begin{equation}\label{sourcecountmodel.eq}
\frac{dN(>S)}{dS}\propto\left[\left(\frac{S}{S_0}\right)^\alpha+\left(\frac{S}{S_0}\right)^\beta\right]^{-1};
\end{equation}
we adopt $S_0$=1.8mJy, $\alpha$=1, $\beta$=3.3 (Scott et al.\ 2002),
although varying these parameters within reasonable ranges
($S_0$=0.5--5, $\alpha$=0.5--2, $\beta$=2--4) does not drastically
affect our results.  We then multiply the assumed source count model
of Equation~\ref{sourcecountmodel.eq} by the transform matrix
$S_{true,obs}$ to obtain the ``observed'' source counts.  The ratios of
the integrated source counts in the underlying source count model to
those ``observed'' by our procedure, give us the correction factors
that need to be applied to the raw source counts in
Table~\ref{FLScounts.tab} to correct for incompleteness and flux bias.

The corrected integrated source counts are given in Column 4 of
Table~\ref{FLScounts.tab} and are plotted in Figure~\ref{counts.fig}
together with counts from other surveys.  Our 850\um\ FLS source
counts are clearly in agreement with both the results of the 8mJy
survey of Scott et al.\ (2002) and with the HDF data of Borys et al.\
(2003) and we conclude that, at least on the basis of source number
densities, there is no evidence that our FLS subfield is not
representative of the sub-mm galaxy population.

\section{SUMMARY AND DISCUSSION}

\begin{figure}
\plotone{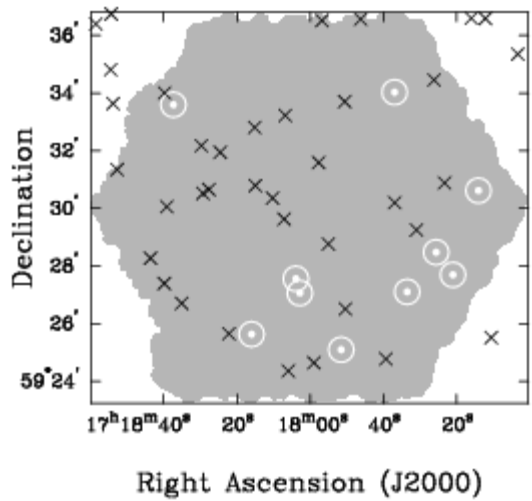}
\caption{\label{radio-posn.fig} Positions of VLA radio detections
compared to our 850\um\ SCUBA sources. Radio detections are shown as
crosses, while SCUBA sources are represented with white circled
points, with the size of the {\it inner point} corresponding roughly
to the 15\arcsec\ diameter of the SCUBA beam.  None of our SCUBA
sources have radio detections. A HIGH QUALITY VERISION OF THIS FIGURE CAN BE OBTAINED FROM THE AUTHORS}
\end{figure}

In this Letter we presented our SCUBA observations of a 151 \sqam\
subarea in the northern CVZ field of the {\it Spitzer} First Look
Survey.  We found a total of 10 sources at S/N$>$3.5 and make their
particulars available to the community. Our integrated source counts
are consistent with those of the other two surveys of the bright end
of the sub-mm population, namely the 8mJy Survey (Scott et al.\ 2002)
and the Hubble Deep Field supermap (Borys et al.\ 2002, 2003).  Given
that extragalactic sub-mm sources cluster strongly on the scales of
current surveys (see, e.g., Figure~\ref{maps.fig}), the fact that our
number counts agree with those of the other two surveys gives an
important confirmation of the numbers of SCUBA sources at the bright
end of the population. Equally significantly, the agreement between
our number counts and those of other surveys suggests that the
subfield of the FLS that we imaged is not unrepresentative of the
extragalactic sky.

We will study the multi-wavelength properties of the sub-mm sources in
our map once the {\it Spitzer} FLS data are released, In the meantime,
we have compared the positions of our SCUBA detections with those of
objects in the 1.4GHz VLA map of the FLS (Condon et al.\ 2003).  There
is no correspondence between the radio-selected and the sub-mm
selected populations (see Figure 3) down to the limit of the VLA
catalog (115 mJy, 5$\sigma$).  This lack of radio detection of any of
our sub-mm sources can be used to constrain their redshifts (Dunne,
Clements \& Eales 2000; see also Yun \& Carilli, 2002): given the VLA
flux limit and the rather narrow dynamic range of these data, most of
our sub-mm sources appear to be at $z$$\gtrsim$ 1.6--1.7 and the
brightest two at $z$ $\gtrsim$ 2.0, (however, these results are likely
to be affected by flux boosting).  These redshift constrains are in
line with our current knowledge of the redshift distribution of sub-mm
selected sources: the median redshift of the population is believed to
lie at $z \sim$ 2--3, and evidence exists of a flux-redshift relation,
such that more luminous sub-mm selected systems (such as those in our
survey) reside at higher redshifts than the less luminous objects
(Ivison et al.\ 2002, Smail et al.\ 2002, Webb et al.\ 2003, Chapman
et al.\ 2003, Clements et al.\ 2004). The lack of 1.4 GHz detection of
our sub-mm sources is thus not unexpected: deeper radio data will be
needed to detect and identify these systems.  We will explore these
issues further once the {\it Spitzer} FLS data are released.

%\acknowledgements 

We thank the Joint Astronomy Centre\notetoeditor{Editor: note that
this is an institution whose name that follows the British spelling
convention of centre/center} staff who helped us obtain these data,
and our colleagues who carried out some of these observations through
the Canadian flexible scheduling scheme.  We also thank Colin Borys
for providing in digital format the number counts shown in
Fig.~\ref{counts.fig}.  The JCMT is operated by the Joint Astronomy
Centre\notetoeditor{Editor: British spelling here, as above} on behalf
of the Particle Physics and Astronoy Reseach Council of the United
Kingdom, the Netherlands Organisation for Scientific Research, and the
National Research Council of Canada.

%\newpage

%% Generally speaking, only the figure captions, and not the figures
%% themselves, are included in electronic manuscript submissions.
%% Use \figcaption to format your figure captions. They should begin on a
%% new page.

\clearpage

\clearpage

%% Tables should be submitted one per page, so put a \clearpage before
%% each one.

%% Two options are available to the author for producing tables:  the
%% deluxetable environment provided by the AASTeX package or the LaTeX
%% table environment.  Use of deluxetable is preferred.
%%

%\input tab1.tex
%\input tab2.tex

%% The following command ends your manusript. LaTeX will ignore any text
%% that appears after it.

\end{document}